\documentstyle[revtex,eqsecnum]{aps}
\begin{document}
\begin{title}
The Role of Vortices in the Mutual Coupling \\
of Superconducting and Normal-Metal Films
\end{title}
\author{
Efrat Shimshoni$^{(a,b)}$
}
\begin{instit}
$^{(a)}$Beckman Institute,
405 North Mathews Avenue,\\
$^{(b)}$Department of Physics,
1110 West Green Street,\\
University of Illinois at Urbana-Champaign,
Urbana, Illinois 61801-3080, USA
\end{instit}

\begin{abstract}

I propose a possible explanation to a recently observed ``cross-talk'' effect
in metal-insulator-metal trilayers, indicating a sharp peak near
a superconducting transition in one of the metal films.
Coulomb interactions are excluded as a dominant coupling mechanism,
and an alternative is suggested, based on the local fluctuating
electric field induced by mobile vortices in the superconducting layer.
This scenario is compatible with the magnitude of the peak signal
and its shape; most importantly, it addresses the
 {\it non-reciprocity} of the effect in exchanging the roles of the films.

\end{abstract}

\newpage
\narrowtext
In a recent experiment \cite{REF:GM} Giordano and Monnier have
observed an intriguing effect in structures composed of two parallel metal
films, separated by a thick insulating layer which prohibits  tunneling.
The voltage measured on one metal film in response to a transport current
in the other exhibited a peak in the narrow interval of temperatures,
$T_c<T<T_{MF}$, in which one of the films (a dirty Al) undergoes a
superconducting-to-normal metal transition (SNT), while the other
(typically Sb) is a normal metal (these films are denoted below by S and N,
respectively). Out of this interval, the induced voltage was negligibly small.
Most astonishingly, the effect was found to be non-reciprocal:
the induced voltage for given $T$ and drive current depended
on which film carries the current. In particular, in the case where current
is driven in S (``case A''), the voltage was a {\it non-linear} function
of the current, and negative (i.e., opposite in sign to the voltage generating
the current); in ``case B'', where the current is driven in N,
the voltage is approximately a linear function of the current, and positive.
The effect was qualitatively the same in the presence of externally applied
magnetic fields (which primarily modify $T_c$ and $T_{MF}$)
- the signal was restricted to the immediate vicinity of the SNT in S.

The purpose of the present work is to propose a coupling mechanism between
superconducting and normal-metal films, which provides a plausible explanation
to most of the observations described above. At first sight, the effect is
reminiscent of the Coulomb drag observed in semiconductor heterostructures
\cite{REF:DRAG} -- \cite{REF:ZM}, which results from Coulomb interactions
between charge carriers across the insulating barrier separating two conducting
layers. As was pointed out in ref. \cite{REF:GM}, in the present case where
{\it metals} are involved Coulomb drag is far too weak to explain their data
at the peak of the signal \cite{REF:metdrag}.
As I show below, a certain enhancement of the drag may
be generated by the presence of vortices in S near the SNT;
however, for the parameters of the system at hand, this enhancement is
insignificant, and I conclude that Coulomb interactions should be ruled out
as a dominant coupling mechanism. I suggest an alternative, which can be
viewed as a coupling between charge carriers in N and vortices
in S, mediated by the flux tubes carried by the latter. The
fact that this mechanism (named  ``inductive coupling'') is dominated by the
dynamics of excitations confined to only one of the layers (i.e.,
vortices), turn out to be a key to understanding the
non-reciprocity of the effect.

A significant clue is the effect being restricted to the region of
SNT in S, and maximized at an intermediate temperature $T_p$,
in which the resistance of that layer is
finite but significantly smaller than the normal state value. It is therefore
natural to suspect (as was also remarked in ref. \cite{REF:GM}),
that the presence of vortices in S is playing a crucial role.
Vortices exist in these thin, dirty Al films since the effective penetration
depth for perpendicular magnetic fields, $\Lambda$, is typically much larger
than the coherence length $\xi_{eff}$ \cite{REF:Pearl}:
\begin{equation}
\Lambda={2\lambda^2_L\over d}\Bigg({\xi_0\over {\ell}}\Bigg)\; ,
\quad \xi_{eff}=(\xi_0\ell)^{1/2}\; ,
\label{EQ:EffL}
\end{equation}
where $d$ is the film thickness, ${\ell}$ is the elastic mean free path,
and $\lambda_L$, $\xi_0$ are, respectively, the intrinsic penetration depth
and $T=0$ coherence length of the material. Free vortices can be
thermally excited near the SNT, even in the absence of externally applied
magnetic field, and the phase slips associated with their motion are known to
be a major source of the finite resistance near the
transition \cite{REF:HN},\cite{REF:KS}.

The strength of coupling between the metallic layers is measured by the
trans-resistivity
\begin{equation}
\rho_{ns}={\cal E}_n/ j_s\, ,
\label{EQ:Rt}
\end{equation}
where ${\cal E}_n$ is the parallel electric field induced in N in response
to a current density $j_s$ in S (adapting the scenario of ``case A'').
It is most useful to relate this transport coefficient (in linear response)
to a correlation function, similarly to the Kubo formula for conductivities
(which, by itself, is less convenient in the
problem at hand). Employing the memory functional formalism \cite{REF:Forst},
\cite{REF:ZM}, one can indeed express the d.c. trans-resistivity (at finite
$T$ and to lowest order in the inter-layer interaction) as
\begin{equation}
\rho_{ns}={1\over k_BTn_nn_se^2A}\int_0^\infty
dt\langle F_n(t)F_s(0)\rangle\, ;
\label{EQ:FFcor}
\end{equation}
here $n_i$ is the electron density in the (two-dimensional) layer $i$,
$A$ is the cross section area of the layers, and
$F_i$ is the time-dependent, zero-wavevector Fourier component of the
force density operator, acting on charge-carriers in layer $i$.
Note the resemblance to the Kubo formula, with the current operators $J_i$
being replaced by their conjugates ${\dot J}_i\propto F_i$. Searching for the
dominant inter-layer coupling mechanism amounts to identifying the components
of $F_i$ dominating the above correlator.

In the rest of the paper, I first consider the Coulomb drag effect
in the vicinity of the SNT, showing that the presence of vortices may enhance
it. This part of the work is concluded by excluding its relevance to the
experiment in ref. \cite{REF:GM}; however, it is an interesting possibility
that may be detectable in different experimental scenarios (e.g., involving
high $T_c$ superconducting films), and to the best of knowledge has not been
proposed elsewhere. In the last and main part of this work I construct
the inductive coupling picture, show that it is compatible with most aspects
of the experiment, and elaborate on the difficulties,
possible resolutions and suggestions for further research.

\noindent
{\bf Enhancement of Coulomb Drag}: ``Coulomb drag'' denotes the finite
trans-resistivity resulting from Coulomb interactions
between charge fluctuations at the different layers. Pictorially, moving
charges in one layer exert a force on charges in the other, thus ``dragging''
them along the direction of the drive current flow. The strength of this
inter-layer coupling indicates the ability of electronic states in the layers
to support inhomogeneities in the charge density, which are necessary to
establish forces between the layers. Substituting the Coulomb force for $F_i$
in Eq.~(\ref{EQ:FFcor}) yields \cite{REF:ZM}
\begin{equation}
\rho_{ns}={\hbar^2\over k_BT\pi n_n
n_se^2}\int{d^2 q \over(2\pi)^2}q^2|V(q)|^2\int_0^\infty d\omega
{{\rm Im}\chi_n(q,\omega)
{\rm Im}\chi_s(q,\omega)\over 4{\rm sinh}^2(\hbar\omega/2k_BT)}\; ,
\label{EQ:RtUD}
\end{equation}
where $V$ is the screened inter-layer Coulomb interaction, and
${\rm Im}\chi_i(q,\omega)$ is the dissipative part of the density-density
response function of layer $i$. The physics described above is reflected by
the $T \rightarrow 0$ behavior of $\rho_{ns}$ being sensitive to the
$\omega\rightarrow 0$, {\em finite} $q$ form
of ${\rm Im}\chi_i(q,\omega)$.

Far below the SNT in layer S, frictional drag is expected to vanish on the
same basis as any dissipation channel, due to the gap to excitations deep in
the superconducting state. Above the transition,
$\rho_{ns}\sim 10^{-6}\, \Omega$
(\cite{REF:GM}-\cite{REF:ZM}) which is negligibly small. However, following
the preceding discussion, the Coulomb drag could in principle be enhanced in
a scenario where density-fluctuations of finite $q$ are favored. I argue
that such a scenario is realized in the close vicinity of the SNT
(just below $T_{MF}$), where a small superconducting gap is opened.
Due to the formation of vortices, this gap is
not uniform -- it vanishes in the core of the vortices, and grows towards their
periphery over a length scale $\xi_{eff}$. Hence, the normal quasi-particle
excitations contributing to ${\rm Im}\chi_s^{(-)}$ (the response function
for $T<T_{MF}$) accumulate at the vortex cores.

I simulate this situation by a simple ansatz for the quasi-particle
eigenstates. Their basis wave-functions, $\psi^-_i({\bf r})$, are
related to the normal-state, $T>T_{MF}$
eigenstates $\psi_i^+({\bf r})$ via an envelope function, which mimics the
spatial variation of the gap in the presence of a dense array of vortices:
\begin{equation}
\psi_i^-({\bf r})=N\psi_i^+({\bf r})[1+
(l_v/\xi_{eff}){\rm cos}({\bf k}\cdot{\bf r})]\; ,
\label{EQ:QPES}
\end{equation}
where $l_v$ is the typical distance between vortices,
$\vert{\bf k}\vert=1/l_v$, and $N\equiv (1+(l_v/\xi_{eff})^2)^{-1/2}$
\cite{REF:deGen}.
Consequently, $\chi_s^{(-)}({\bf q},\omega)$ \cite{REF:VolWol} can be cast
in a form involving $\chi_s^{(+)}$ (of $T>T_{MF}$)
with ${\bf q}$ shifted by $\pm{\bf k}$.
When substituted in Eq.~(\ref{EQ:RtUD}), these terms
involving $\chi_s^{(+)}({\bf q\pm k},\omega)$ will dominate, due to the pole
established at $\omega=0$, ${\bf q=\pm k}$. Assuming further that the normal
state (of both metallic layers, S and N)
is in the diffusive regime, so that $\chi_s^{(+)}({\bf q},\omega)
=(dn/d\mu)Dq^2/(Dq^2-i\omega)$ (with $dn/d\mu$ the density of states
and $D$ the diffusion coefficient), I obtain an approximate
expression for $\rho_{ns}^{(-)}$,
\begin{equation}
\rho_{ns}^{(-)}\sim\rho_{ns}^{(+)}(\hbar D/l_v^2k_BT)^{1/2}
(l_v/\xi_{eff})^4\; .
\label{EQ:RtCD}
\end{equation}
The first factor in parenthasis tends to {\em enhance} the drag, in comparison
with its value for $T>T_{MF}$, reflecting the intuition that non-uniform
density fluctuations are stabilized by the non-uniform gap. However, this
competes with the second, suppressing factor, associated with the amplitude
of the gap modulation. Near the SNT, the former increases with $T$ (with
increasing vortex density), while the latter decreases due to the divergence of
$\xi_{eff}(T)$. On the face of it, this behavior is qualitatively compatible
with the observation in ref. \cite{REF:GM}. However, the maximum of
$\rho_{ns}^{(-)}$ is achieved at $l_v\sim\xi_{eff}$, in which case it is
enhanced with respect to $\rho_{ns}^{(+)}$ by at most an order of magnitude,
for the parameters of the system at hand. As  $\rho_{ns}^{(+)}$ is
extremely small in the first place, one must conclude that this enhancement
of Coulomb drag is irrelevant to the present experiment. A more pronounced
enhancement may be expected, however, if the Al is replaced by a material
with higher $T_c$, shorter $\xi_{eff}$ and lower $D$ (e.g., near a
superconducting-insulator transition), and could serve as an interesting
demonstration of coupling between density fluctuations and phase-fluctuations
of the superconducting order parameter.

\noindent
{\bf Inductive Coupling}: An alternative coupling mechanism between the
layers is associated with the motion of free vortex and anti-vortex
excitations in S at $T_c<T<T_{MF}$. These excitations are accompanied by
(self-consistently generated)
magnetic flux tubes, which extend out perpendicularly to S
and penetrate the neighboring layer N. Note that the magnetic field
of a vortex varies over length scale $\Lambda\sim 25\mu{\rm m}$ (for
$\lambda=500 \AA$, $\xi_0=1.6\mu{\rm m}$, $\ell\sim 100 \AA$ and
$d\sim 350 \AA$ in Eq.~(\ref{EQ:EffL})), that is much larger than the
thickness of the trilayer device, so that bending of the field lines
out of S can be neglected. In the absence of an external
magnetic field, the fluctuating magnetic field thus generated in N,
${\bf B}({\bf r},t)$, averages to zero over the sample area, as vortex and
anti-vortex excitations are equally likely. However, the local,
instantaneous time-dependence
of ${\bf B}({\bf r},t)$ induces a fluctuating electric field
${\bf E}_n({\bf r},t)$. Since ${\bf E}_n({\bf r},t)$ is correlated with the
electric field in S, a finite inter-layer coupling coefficient is established.

I first focus on ``case A'', in which the drive current is passed in the
layer S. To evaluate the trans-resistivity $\rho_{ns}$ using
Eq.~(\ref{EQ:FFcor}), I consider the force fluctuation ${\bf F}_s({\bf r},t)$
acting on the charge carriers in S due to the phase-slips associated with
vortex motion \cite{REF:HN}:
\begin{equation}
{\bf F}_s({\bf r},t)=(\phi_0/c){\bf{\hat z}\times J}_v({\bf r},t)\; ,
\label{EQ:Fs}
\end{equation}
where $\phi_0=hc/2e$ is the flux quantum, ${\bf{\hat z}}$ is a unit vector
perpendicular to the layer, and ${\bf J}_v({\bf r},t)=n_v({\bf r},t)
{\bf v}_v({\bf r},t)$ is the fluctuating {\em vortex}-current density
($n_v({\bf r},t)$ and ${\bf v}_v({\bf r},t)$ being the vortex density and
velocity, respectively). The force in N is, in turn,
${\bf F}_n({\bf r},t)=e{\bf E}_n({\bf r},t)$, where ${\bf E}_n$ satisfies
\begin{equation}
\nabla\times {\bf E}_n({\bf r},t)=-(1/c)\partial {\bf B}({\bf r},t)/
\partial t\; .
\label{EQ:MaxEq}
\end{equation}
${\bf B}({\bf r},t)$ is assumed the form
\begin{equation}
{\bf B}({\bf r},t)=\phi n_v({\bf r},t)\; ,
\label{EQ:FlucB}
\end{equation}
where $\phi$ is an effective flux transferred by a single moving vortex. The
distance that a vortex can traverse freely is limited by $l_v$, the typical
inter-vortex spacing (beyond which it is likely to be annihilated by an
anti-vortex). In the present case, near the center of the SNT region
$l_v\ll\Lambda$, and hence the flux-transfer generated in N along with a
phase-slip of $2\pi$ in S {\em is much smaller than $\phi_0$}.
Approximating the magnetic field within a radius $\Lambda$ of a vortex
by its average, and using a crude ratio-of-areas argument, I find
\begin{equation}
\phi= f\phi_0\; ,\quad f\sim l_v/2\Lambda\; .
\label{EQ:PhiEff}
\end{equation}
Eq.~(\ref{EQ:MaxEq}) Combined with
Eqs.~(\ref{EQ:FlucB}),  ~(\ref{EQ:PhiEff}) yields (ignoring fast fluctuations
with $\nabla\cdot v_v\not=0$ \cite{REF:HN})
\begin{equation}
{\bf F}_n({\bf r},t)=(\phi/c){\bf{\hat z}\times J}_v({\bf r},t)\; .
\label{EQ:Fn}
\end{equation}
Comparing to Eq.~(\ref{EQ:Fs}), one observes that ${\bf F}_n$ differs from
${\bf F}_s$ only by the reduction factor $f$ relating $\phi$ to $\phi_0$,
reflecting the fact that it is induced by the ``magnetic fraction'' of the
very same vortices. Using Eq.~(\ref{EQ:FFcor}), I thus obtain
\begin{equation}
\rho_{ns}={1\over k_BT}\Bigg({\phi\phi_0\over c^2}\Bigg)\int_0^\infty
dt\langle J_v(t)J_v(0)\rangle=\Bigg({\phi\phi_0\over c^2}\Bigg)\sigma_v
\label{EQ:JJcor}
\end{equation}
($J_v(t)$ is the $q=0$ Fourier component of ${\bf J}_v({\bf r},t)$, along the
direction of the drive current); the last equality follows from a ``Kubo
formula'' for the vortex-conductivity $\sigma_v$. I next assume that the
resistivity of S near $T_p$ is dominated by the vortex flow, i.e.
$\rho_s=(\phi_0/c)^2\sigma_v$ \cite{REF:HN}. This implies that, within the
assumptions above,
\begin{equation}
\rho_{ns}(T)=f(T)\rho_s(T)\; .
\label{EQ:RtfR}
\end{equation}
Since $\rho_s(T)$ increases as a function of $T$, while $f(T)$ decreases
(See Eq.~(\ref{EQ:PhiEff}), noting that $l_v$ decreases and, at the same time,
$\Lambda$ diverges), $\rho_{ns}(T)$ is non-monotonic for $T_c<T<T_{MF}$,
and vanishes for $T<T_c$ (where $\rho_s=0$) and $T>T_{MF}$ (where $f=0$). This
is in qualitative agreement with the experimental result.

So far I relied on the assumption of {\em linear response}, which is in fact
inconsistent with the experiment. Before discussing the consequence of
relaxing this assumption, note that within linear response, the Onsager
relations would imply reciprocity, namely $\rho_{ns}=\rho_{sn}$ (where
$\rho_{sn}$ is the transport coefficient compatible with ``case B''). To
confirm this, I focus on ``case B'', in which a current density $j_n$
is passed in N. This current, enforced by the external driving source, applies
a Lorentz force on the flux tubes carried by vortices in S \cite{REF:KS},
and thus on the vortices themselves:
\begin{equation}
{\bf F}_L=(\phi/c){\bf{\hat z}\times j}_n\; ,
\label{EQ:Flor}
\end{equation}
where $\phi$ is given by Eq.~(\ref{EQ:PhiEff}). ${\bf F}_L$ resembles the
force on a vortex in the presence of a supercurrent \cite{REF:HN}, except that
the ``vortex charge'' is effectively renormalized by the fraction $f$.
An equivalent scenario, in terms of the impact on the vortices,
could be achieved by driving a current $j_s=fj_n$ {\em directly through S}.
In view of this equivalence, I find, as expected,
\begin{equation}
\rho_{sn}=f\rho_s=\rho_{ns}\; .
\label{EQ:RtSN}
\end{equation}

The violation of reciprocity observed in ref. \cite{REF:GM} can indeed be
associated with {\em non-linearity},
and the argument is as follows: in ``case A'',
the applied supercurrents $j$ are assumed sufficiently strong to push the
vortices into the non-linear response regime. The principal implication on
Eq.~(\ref{EQ:JJcor}) is that the vortex response is replaced by a
current-dependent coefficient, $\sigma_v(j)$. However, in ``case B'' where
{\em the same} current $j$ is supplied to N, the vortices respond as if a
much smaller current $fj$ is driven directly in S; their conductivity is
hence well-approximated by $\sigma_v(0)$. This distinction between the two
cases is clearly consistent with the experiment.

The estimated magnitude of $\rho_{sn}$ at the peak $T_p$ is also in accord
with the data presented in ref. \cite{REF:GM}. I evaluate $f$ using the
Landau-Ginzburg expression $\Lambda(T_p)=\Lambda(0)(T_{MF}/(T_p-T_{MF}))$
(see Eq.~(\ref{EQ:EffL})), and the analysis of ref. \cite{REF:HN} to
estimate $l_v(T_p)$.
For the experimental values $T_c=1.77\, K$, $T_p=1.81\, K$
and $T_{MF}=1.86\, K$, I get $f\sim 10^{-4}$, which implies a peak voltage
$V_s\sim 200\, {\rm nV}$ in S (with film resistance
$R_s(T_p)\sim 400\, \Omega$) for a current $I=7\, {\rm \mu A}$ driven in N.
This is roughly a factor of 2.5 {\em larger} than the experimental value. A
better quantitative agreement can not be expected within the
crude assumptions involved in this work. In particular, my
over-estimate of the effect may be due to over-estimating the contribution
of vortex dynamics to $R_s$, disregarding other degrees of freedom.

Finally, the sign of the effect in the linear case (B) can be reconciled with
the mechanism proposed here: the arguments leading to Eq.~(\ref{EQ:RtSN})
also imply that it is identical to the sign of a voltage established in S
in response to an `equivalent', in-layer supercurrent. The negative
sign in case A is the one aspect of the experiment still open for
interpretation. It is not, however, in contradiction with the picture
constructed so far: since it occurs in the non-linear response regime
\cite{REF:signrev}, there is a good reason to suspect that it involves
processes not included in the present scheme (which, e.g.,
introduce a force on the vortices in the opposite direction). Most likely are
thermal conduction processes, leading to flow dictated by a temperature
gradient rather than electric current \cite{REF:Therm}. A more
elaborate theory is required to clear this point. Further experimental work
could also shed some light, e.g. a more detailed study of the
dependence on drive current. It should be noted that the inductive coupling
mechanism can be distinguished from Coulomb drag by a multitude of tests:
the latter is more sensitive to the distance between layers and to the sign
of charge carriers; moreover, for a narrower strip \cite{REF:NotrDam} Coulomb
interaction is expected to be enhanced, while induction is further limited
by the width of the strip. I conclude by pointing out that in view of the
interpretation proposed in this paper, the ``cross-talk'' effect is a
suggestive probing technique for the dynamical properties of vortices (in
distinction from other degrees of freedom in a superconducting film).
\acknowledgements
I am grateful to N.\ Giordano for showing me his data prior to publication,
and to him as well as S.\  Girvin, A.\  Leggett and S.\ Sondhi for
illuminating discussions. This work was supported by the Beckman Foundation.



\begin{references}
\bibitem{REF:metdrag}
In comparison with semiconductors, the Fermi energy is larger, and in addition
screening is more effective.
\bibitem{REF:signrev}
For sufficiently small drive currents one expects the sign of the signal in A
to be eventually reversed so as to coincide with case B in the linear
response regime. However, the signal in that case would presumably be too
small.
\end{references}
\end{document}